\documentclass[letterpaper]{article} 
\usepackage{aaai23}  
\usepackage{times}  
\usepackage{helvet}  
\usepackage{courier}  
\usepackage[hyphens]{url}  
\usepackage{graphicx} 
\usepackage{natbib}  
\usepackage{caption} 
\usepackage{algorithm}
\usepackage{algorithmic}

\usepackage{newfloat}
\usepackage{listings}
\DeclareCaptionStyle{ruled}{labelfont=normalfont,labelsep=colon,strut=off} 
\floatstyle{ruled}
\newfloat{listing}{tb}{lst}{}
\floatname{listing}{Listing}
\usepackage[labelsep=period]{caption}

\title{Popular Support for Balancing Equity and Efficiency in Resource Allocation:\\
A Case Study in Online Advertising to Increase Welfare Program Awareness}
\author {
    Allison Koenecke\textsuperscript{\rm 1},
    Eric Giannella\textsuperscript{\rm 2},
    Robb Willer\textsuperscript{\rm 3},
    Sharad Goel\textsuperscript{\rm 4} \\
}
\affiliations {
    \textsuperscript{\rm 1} Cornell University \\
    \textsuperscript{\rm 2} Code for America \\
    \textsuperscript{\rm 3} Stanford University \\
    \textsuperscript{\rm 4} Harvard Kennedy School \\
    koenecke@cornell.edu, 
    eric@codeforamerica.org, 
    willer@stanford.edu,
    sgoel@hks.harvard.edu
}

\begin{document}

\maketitle

\begin{abstract}
Algorithmically optimizing the provision of limited resources is commonplace across domains from healthcare to lending. 
Optimization can lead to efficient resource allocation, but, if deployed without additional scrutiny, can also exacerbate inequality.
Little is known about popular preferences regarding acceptable efficiency-equity trade-offs, making it difficult to design algorithms that are responsive to community needs and desires.  
Here we examine this trade-off and concomitant preferences in the context of
GetCalFresh, an online service that streamlines the application process for California’s Supplementary Nutrition Assistance Program (SNAP, formerly known as food stamps). GetCalFresh runs online advertisements to raise awareness of their multilingual SNAP application service.  We first demonstrate that when ads are optimized to garner the most enrollments per dollar, a disproportionately small number of Spanish speakers enroll due to relatively higher costs of non-English language advertising. Embedding these results in a survey  (N = 1,532) of a diverse set of Americans, we find broad popular support for valuing equity in addition to efficiency: respondents generally preferred reducing total enrollments to facilitate increased enrollment of Spanish speakers. These results buttress recent calls to reevaluate the efficiency-centric paradigm popular in algorithmic resource allocation.
\end{abstract}

\section{Introduction}

In allocating limited resources, managers and policymakers now regularly turn to statistical algorithms designed to achieve efficient outcomes. 
For example, in healthcare systems around the country, statistical algorithms are routinely used to target patients for ``high-risk care management,'' a costly benefit made available only to those with the most pressing healthcare needs~\cite{obermeyer2019dissecting}.
Child services agencies use statistical models to identify children at risk of maltreatment, prioritizing them for follow-up visits and potential interventions~\cite{chouldechova2018case,de2020case,shroff2017predictive}.
And, building inspectors in New York City are guided by an algorithm designed to identify structures that pose the greatest safety risk~\cite{flowers}, making efficient use of scarce inspection resources.

In these examples and others, optimization can be a boon to communities, efficiently allocating scarce resources in a manner that is designed to save lives. 
But with the rise of statistical optimization, there has been growing recognition that algorithms themselves can entrench and exacerbate inequality. 
For example, to determine which patients to prioritize for high-risk care management, one popular algorithm predicted healthcare costs as a proxy of medical need; due to unequal access to medical resources, Black patients in the training dataset incurred lower medical costs than equally sick white patients, in turn meaning that Black patients were disproportionately de-prioritized by the algorithm for valuable medical treatment~\cite{obermeyer2019dissecting}.

To understand and help ensure the equity of algorithmically guided decisions, the fair machine learning literature has, at a high-level, focused on three broad and complementary approaches. 
First, researchers have worked to make predictive algorithms as accurate as possible for all subgroups of a population~\cite{hebert2018multicalibration}---and have highlighted the subtle ways in which seemingly reasonable approaches can lead to poor performance, as with the healthcare algorithm discussed above~\cite{buolamwini2018gender,koenecke2020racial}.
Second, an extensive body of literature aims to articulate axioms for ensuring that algorithms are equitable---like requiring error rates be equal across subgroups of a population---and to develop corresponding procedures that ensure algorithms adhere to these  principles~\cite{darlington1971another,cleary1968test,zafar2017parity, dwork2012fairness,chouldechova2017fair,hardt2016equality,kleinberg2016inherent,woodworth2017learning,zafar2017fairness,corbett2017algorithmic,chouldechova2020snapshot,berk2021fairness}.
Third, and in contrast to the axiomatic approach, a recent strand in the fairness literature argues for focusing on consequences over process, advocating that algorithms be designed to achieve outcomes favored by communities~\cite{L2BF,nilforoshan2022causal}. 
This last approach is motivated in part by the observation that popular fairness axioms can in some important cases lead to worse outcomes for historically marginalized communities~\cite{nilforoshan2022causal}.

One particularly strong appeal of the outcome-focused approach is that it is responsive to community needs and desires.
This advantage, however, is also a limitation, as the approach by design requires input from community members to determine which outcomes are in fact preferred. Furthermore, given the heterogeneity of political and social attitudes, it is far from clear that one can find common ground.

Here we take an initial step at investigating the viability of an outcome-based approach to resource allocation and algorithm design. We do so in the context of the government-run Supplementary Nutrition and Assistance Program (SNAP)---formerly known as food stamps---which helps low-income people buy nutritious food~\cite{cbpp2021}.
We first show through a series of experiments that it is more expensive to run informative online ads regarding SNAP targeting Spanish speakers relative to English speakers.
As these online ads are one of the central ways for enrolling eligible participants into the program, this cost differential creates an inherent trade-off: spending a limited advertising budget to recruit Spanish speakers means fewer total individuals hear about, and ultimately enroll in, the SNAP program.

We then present this trade-off to a diverse sample of Americans, and elicit respondents' preferences by having them select their preferred advertisting strategies in a series of pairwise comparisons.
Our work thus fits within the larger literature on preferences for the distribution of valued resources across large populations (e.g.,~\citet{tversky1981evidential,Norton2011}), and specifically the allocation of resources across groups that may face different degrees of disadvantage.
But, the methods and results stand in contrast to much prior work on moral decision-making that focuses on preferences in dichotomous decision settings impacting one or two individuals (e.g.,~\citet{Awad2018,Hannan2021}).  
Importantly, when presented with population-level options---such as our advertising strategies---respondents can reveal preferences for parity in allocations between groups that are hard or impossible to express when asked about individuals.

Our results show broad popular support for incorporating equity in addition to efficiency in allocation decisions: across groups defined by age, gender, race, and political affiliation, respondents generally preferred reducing total SNAP enrollments to facilitate increased enrollment of Spanish speakers. These findings can be used by advertisers---such as GetCalFresh---to determine fair budget allocations for subgroup-targeting ads.
Importantly, whereas most---and perhaps nearly all---deployed algorithms are designed with a near-exclusive focus on efficiency, our results suggest that, at least in some circumstances, most Americans support reducing efficiency in order to allocate more resources to marginalized groups.

\section{An Equity-Efficiency Trade-Off\\In Online Advertising for SNAP}

Roughly 12\% of Americans annually participate in SNAP~\cite{usda_2020}, 
though many more low-income individuals are eligible.
One barrier to participation is that the sign-up process can be confusing and time-intensive. The online service GetCalFresh\footnote{This research is conducted in partnership with Code for America, a non-profit organization that built and runs GetCalFresh.} aims to address this obstacle by simplifying SNAP enrollment.
To help build awareness of its service, GetCalFresh primarily recruits individuals through online ads. In the United States, the advent of mobile technologies has been an equalizer on this front---about 71\% of adults earning less than \$30,000 per year own a smartphone~\cite{pew2019}, and marketing firms estimate that the average American views 4,000 to 10,000 ads per day~\cite{forbes2017}.  As such, online advertising is a powerful tool for GetCalFresh; however, it is critical that this recruitment strategy does not leave out underserved populations.

\begin{figure}[t]
    \centering
    \includegraphics[width=.9\linewidth]{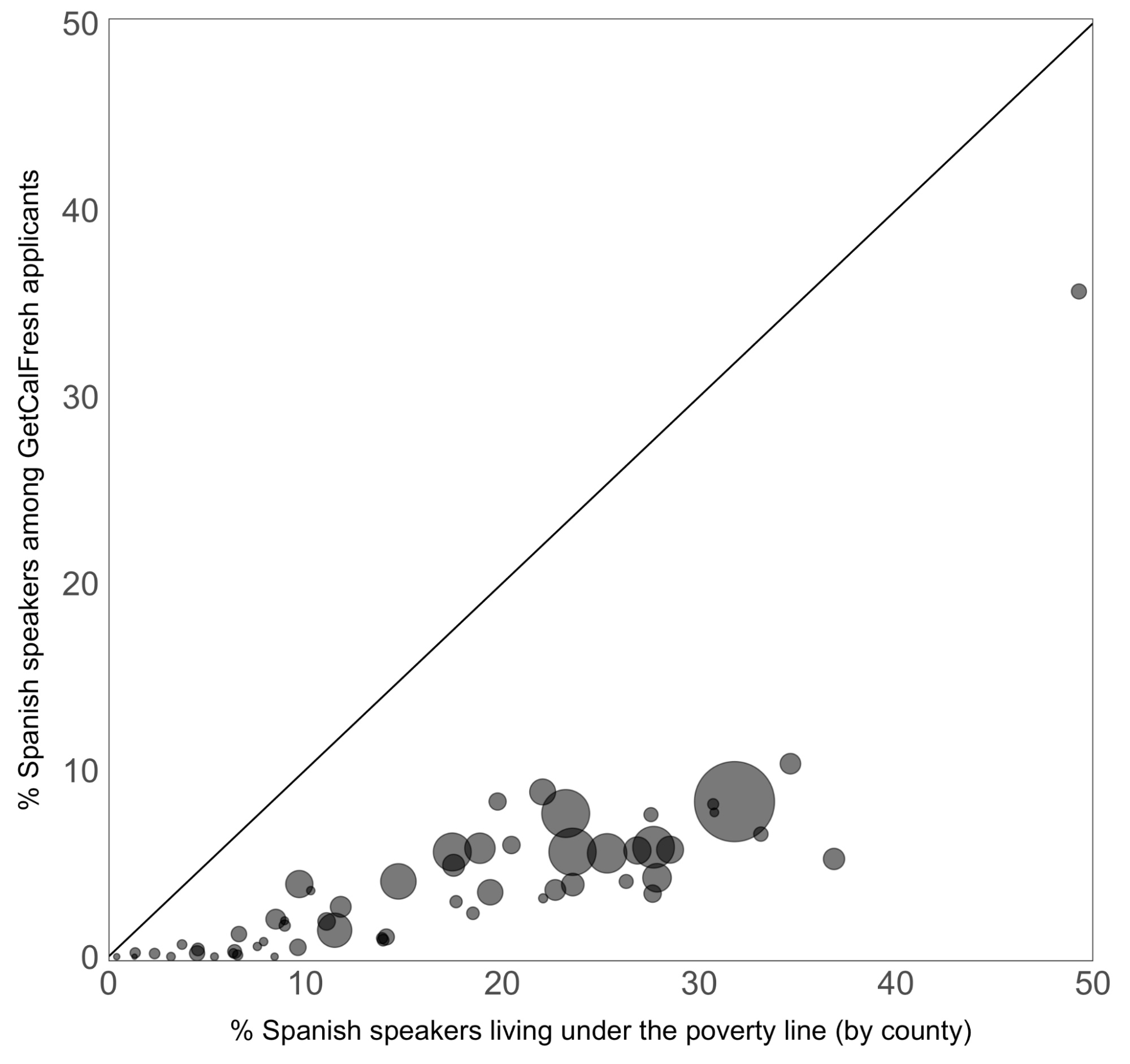}
    \caption{In all California counties, the share of primarily Spanish-speaking individuals is smaller (i.e., below the diagonal) among GetCalFresh applicants relative to the population living under the poverty line---a common proxy for SNAP eligibility.  Scatter dot sizes correspond to total population of county.  In San Diego county, roughly 23\% of adults living below the poverty line primarily speak Spanish at home.}
    \label{fig:county_compare}
\end{figure}

One of GetCalFresh's goals is to ensure that its SNAP enrollees include a ``fair'' share of Spanish speakers. 
We focus on Spanish speakers as a demographic of interest for several reasons.  First, individuals speaking English as a second language often face greater hurdles in accessing government financial support: many instructions are complicated and default to English, and there is confusion and fear about how SNAP usage might affect eligible residents who are not U.S.\ citizens.  Second, GetCalFresh  offers their online SNAP application in both English and Spanish language options, and seeks to publicize to Spanish-speakers that they are also able to use the website.\footnote{This paper will refer to only ``English speakers'' and ``Spanish speakers'' based on the individuals' language preference when filling out a SNAP application (since a language must be explicitly chosen in the form); we recognize that this elides bilingualism and readers of other languages (such as Chinese, which is supported by GetCalFresh but has a low volume of applicants), and discuss implications in the Methods section.}  Third, there are disproportionately few Spanish speakers among GetCalFresh's enrollee demographics relative to the share of Spanish speaking individuals at or below the poverty line (see Figure \ref{fig:county_compare})---living under the poverty line is a good proxy for SNAP enrollment eligibility~\cite{acs2018,moon2019}.

\begin{figure*}[t]
    \centering
    \includegraphics[width=.9\linewidth]{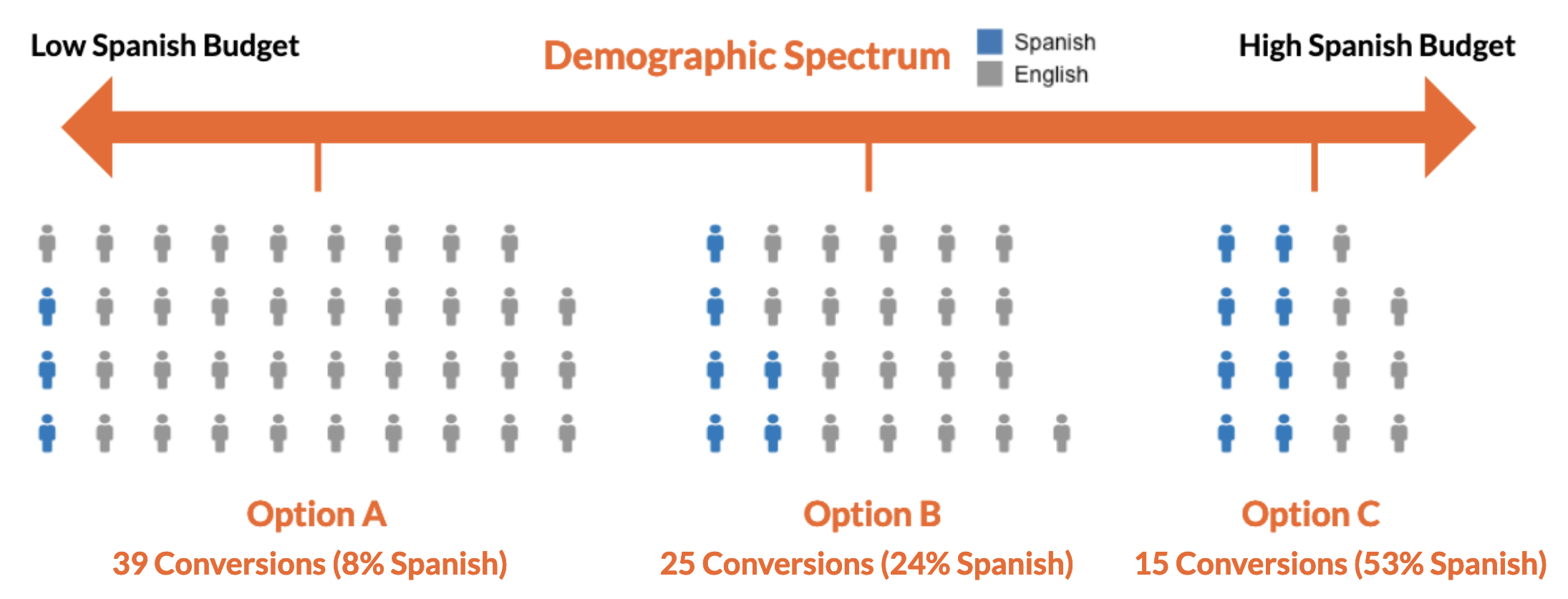}
    \caption{The three options displayed each represents a feasible allocation of the daily advertising budget.  Options A, B, and C are each attainable by spending the same dollar amount, and each yields a different number of total ``conversions'' (i.e., the number of individuals who are shown a GetCalFresh ad and then proceed to fill out a SNAP application via GetCalFresh).  The trade-offs among different options occur because it costs more to target Spanish speakers in online advertising---so, if we increase the Spanish-targeting share of the budget, we will obtain fewer conversions overall. On the left side of this spectrum (Option A): if a low share of the budget goes towards Spanish-speaking individuals, then there will be disproportionately more English-speaker conversions. On the right side of this spectrum (Option C): if a high share of the budget goes towards Spanish-speaking individuals, then there will be disproportionately fewer English-speaker conversions.}
    \label{fig:fairness_compare}
\end{figure*}

GetCalFresh uses online advertising for outreach to both English speakers and Spanish speakers.  When an individual performs a Google search on certain keywords (e.g., ``how to apply for food stamps''), they may be shown a GetCalFresh ad; if they then click on the ad and proceed to fill out and submit the SNAP application, they are counted as a ``conversion.''  When running online ad campaigns, one must consider both the ad-targeted individuals' ensuing conversion rates and the demographic composition of the resulting ``converted'' enrollees.
On the one hand, optimizing exclusively for conversions per dollar---a common measure of utility in online advertising---can lead to an unacceptable demographic distribution, such as one in which all enrollees belong to a single group. This concern is real, as the cost-per-conversion often varies significantly across demographic groups~\cite{lambrecht2016}, meaning that some groups may be inadvertently left behind; this concept is referred to as ``crowding out'' of the market.
On the other hand, imposing strict demographic parity (e.g., requiring that the demographic distribution of recruited individuals matches the composition of the SNAP-eligible population) can result in unacceptably high cost, meaning fewer people overall are ultimately enrolled~\cite{gelauff2020, nasr2020}.  

We think of maximizing conversions as the ``efficient'' strategy.
In our application, Spanish speakers are more expensive to reach via ads (due to the supply-and-demand forces leading to ``crowding out,'' which is quantified in experiments described in the Methods section).  Because GetCalFresh has a limited advertising budget, allocating this budget such that one additional Spanish speaker is converted necessitates counterfactually disallowing the conversion of multiple English speakers. This tension is summarized in Figure \ref{fig:fairness_compare}. 
Among the available advertising options---ranging from maximizing conversions to prioritizing Spanish speakers to various degrees---the key question is then which we should choose.

\begin{figure}[t]
    \centering
    \includegraphics[width=\linewidth]{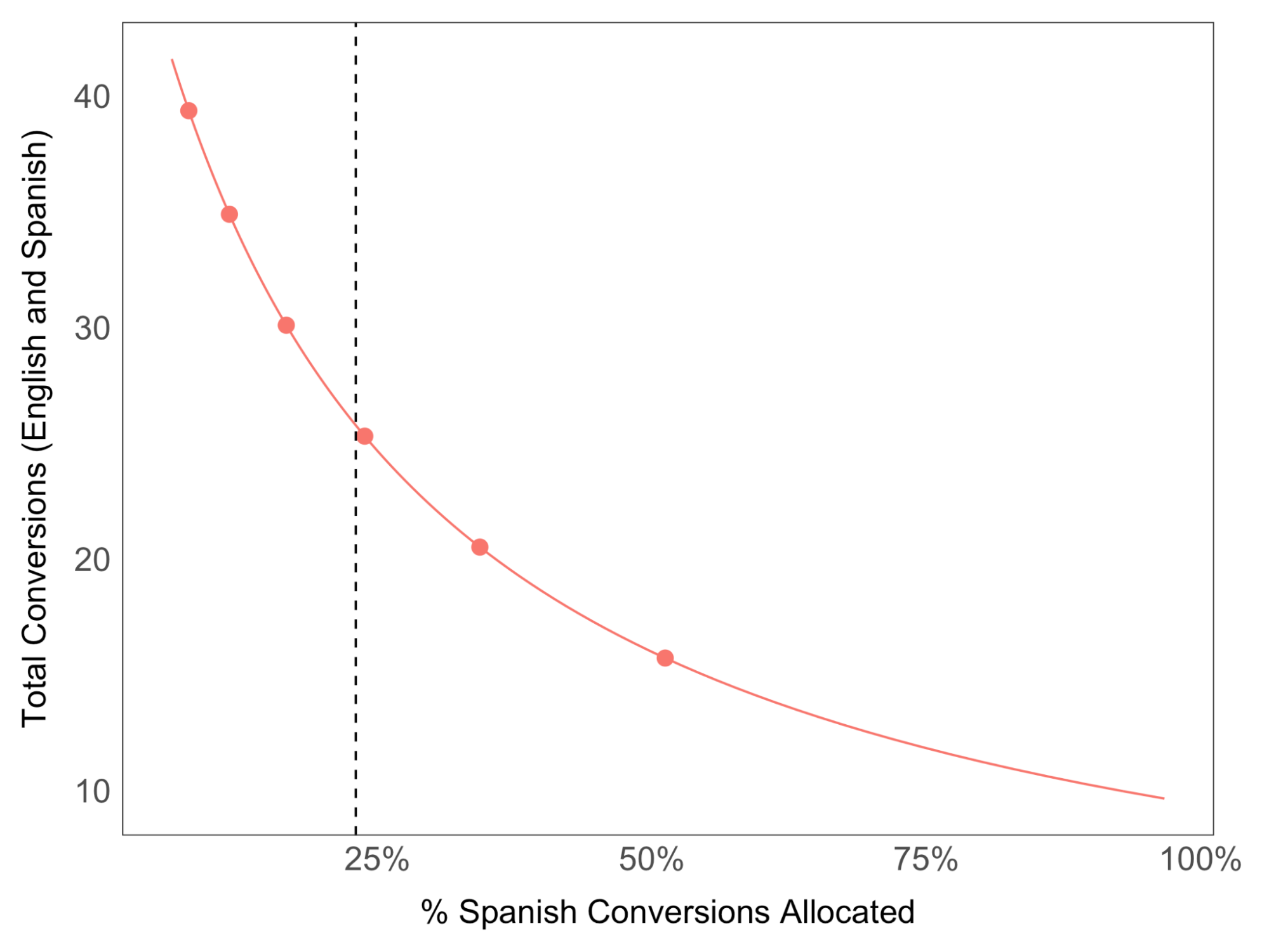}
    \caption{This ``high trade-off'' Pareto frontier is generated based on observational data of the cost-per-conversion of English and Spanish language ad campaigns run in San Diego County. Using a default algorithmic ad bidding optimization for each campaign, we find that the same cost to gain one Spanish conversion could be used to instead gain six English conversions.  The six points plotted along the Pareto frontier represent six feasible budget allocations that we elicit human preferences among; the leftmost and rightmost points here correspond to Options A and C, respectively, in Figure \ref{fig:fairness_compare}.
    The vertical dotted line demarcates that 23\% of adults under the poverty line in San Diego County primarily speak Spanish, which can be used as a benchmark for the ``demographic parity'' definition of fairness.}
    \label{fig:high_pareto}
\end{figure}

To study this ``efficiency-equity'' trade-off, we use a two-step methodology. First, we estimate the expected outcomes for feasible algorithmic advertising policies. 
To do this, we run a series of online ad experiments to estimate the demographic composition of enrollees resulting from different budget-constrained advertising strategies.
The results of these experiments allow us to construct a Pareto frontier that traces out the largest number of conversions we can accrue for a given demographic composition, subject to our budget constraint. 
Second, we elicit preferences over the Pareto-optimal strategies~\cite{tesauro1989, zintgraf2018} to determine how individuals balance competing objectives in this setting.

There is a robust body of work quantifying demographic disparities in Google Ads~\cite{lambrecht2016, lambrecht2020, sweeney2013, datta2015} and other advertising platforms~\cite{ali2019, imana2021, jansen2013}. 
However, relatively little is known about how individuals
balance efficiency-equity trade-offs when presented with a concrete instance of these disparities. 

\section*{Methods}

\subsection*{Google Ad Bidding}\label{sec:google_targeting}
There are several mechanisms that advertisers can use via the Google Ads platform.  First, the advertiser can choose the type of ad to bid on: display ads (shown on websites regardless of the individual's searches), and search ads (based explicitly on Google search keywords).  We focus on search ads because they more directly target potential SNAP enrollees, and are cheaper and more effective than display ads.  Next, the advertiser has a choice of two default machine learning-based methods to bid on users in the ad auction run by Google.  First, one can use the ``Target Cost-Per-Acquisition (CPA)'' method, in which Google auto-bids on users so that the average monthly cost-per-conversion is close to the target dollar value set by the advertiser.  Second, one can use the ``Maximize Conversions'' method, in which the advertiser provides Google with a pipeline for identifying specific types of users who have successfully converted in the past, and Google bids in a way that maximizes the number of conversions similar to these historical users (up to a set monthly budgeted dollar amount).

To more precisely target SNAP-eligible enrollees, we focus on search ads based on a particular set of keywords used to commonly search for SNAP applications.  We generate the most common set of English keywords used historically on Google Ads that led to conversions (e.g., ``apply for food stamps'' or ``sign up for calfresh''), and translate those searches into Spanish keywords (e.g., ``solicitar cupones de alimentos'' or ``registrarse para calfresh'').  We do the same for text-based ad copy; one of three sets of ad text are shown to the user based on their keyword searches---that is, ads regarding one of CalFresh, EBT, or food stamps are shown, either in Spanish or English as described below.\footnote{Keyword translations, survey text, and figure reproduction code are documented at \url{https://github.com/koenecke/equity-efficiency-balance}.}

Using these keyword lists, we run two separate ad targeting campaigns by language.  The English-targeting Google Ad campaign is triggered when a user searches one of the keywords on the list of English language keywords we provide, for a user geographically in a specific county.  Meanwhile, the Spanish-targeting Google Ad campaign is triggered when a user searches one of the keywords on the list of Spanish language keywords we provide, and the user has their Google language settings set to Spanish, for a user geographically in the same specific county.  These two ad targeting campaigns are then run using both ``Target CPA'' and ``Maximize Conversions'' bidding methods.

Focusing on San Diego County, we experimentally observe significantly higher cost-per-conversion for Spanish language SNAP ads relative to English language SNAP ads. When using the ``Maximize Conversions'' bidding method, the average daily cost per conversions for a Spanish-targeting ad campaign is on average 3.8 times more than that for an English-targeting ad campaign.  Even when using the ``Target CPA'' method (which is less prohibitively expensive than the ``Maximize Conversions'' method), we find on average 1.4 times greater daily cost-per-conversions for Spanish-targeting ad campaigns. The reason for these language-based disparities could stem from the demand side: fewer Spanish speakers are Googling for these keywords relative to English-speaking counterparts.  There are other potential reasons for disparities (though we cannot confirm any of these experimentally given lack of internal data access): for example, mis-calibration of quality scores for ads of different languages, or bias in reserve price setting if no other advertisers are competing on these search terms.

The following two observed ``trade-offs'' in San Diego County form the basis for our main analysis:
\begin{enumerate}
    \item \textbf{``Low trade-off''}: The cost to advertise to one Spanish speaker is the same as the cost to advertise to three English speakers using Google's ``Target CPA'' bidding method.
    \item \textbf{``High trade-off''}: The cost to advertise to one Spanish speaker is the same as the cost to advertise to six English speakers using Google's ``Maximize Conversions'' bidding method.
\end{enumerate}

In addition to the ``high trade-off'' and ``low trade-off'' settings, we also study three synthetic trade-offs attainable with the same cost: (a) one Spanish speaker to one English speaker; (b) three Spanish speakers to one English speaker; and (c) six Spanish speakers to one English speaker.  Our main results extrapolate to these synthetic arms as well; detailed results are presented in the Appendix.

\subsection*{Generating Pareto-Optimal Advertising Strategies}
GetCalFresh budgets roughly \$400 daily in Google advertising to anyone searching for food stamps in San Diego, California.  Currently, nearly half a million individuals in San Diego are living under the poverty line; of those, 23\% primarily speak Spanish~\cite{acs2018}.  Meanwhile, the default Google Ads budget allocations resulted in roughly 7\% of conversions attributable to Spanish speakers.  
We can increase this share of Spanish speakers by allocating a larger proportion of the budget to Spanish ads.
However, because Spanish ads cost more per conversion, recruiting more Spanish speakers means recruiting fewer total people, given the fixed budget. 

We depict this trade-off by tracing out the 
corresponding Pareto frontier in Figure \ref{fig:high_pareto}: given our fixed advertising budget, the curve shows the maximum number of conversions attainable for different proportions of Spanish speakers converted.
We derive the frontier for the ``high trade-off'' setting in the figure as follows. If we let 100\% of the daily Google ad budget be spent towards the English-speaking ad campaign, we can average 36 English-speaker conversions and 3 Spanish-speaker conversions.  
In this 100\% English advertising strategy, we expect to recruit some Spanish speakers because some people who prefer to fill out a government form in Spanish may still set their browser language to English or search for the term ``food stamps'' in English.  
Meanwhile, if we let 100\% of the daily Google ad budget be spent towards the Spanish-speaking ad campaign, we can average 7 English-speaker conversions and 13 Spanish-speaker conversions.  
Again, we expect to convert some English-speaking individuals despite running a Spanish targeting campaign, in part because individuals may be bilingual or may be using a shared household computer.

Finally, for any mix of English and Spanish advertising strategies, the resulting demographic composition is a weighted combination of the two extreme options.
For example, suppose we spend 80\% of the daily ad budget on the English targeting campaign and 20\% on the Spanish targeting campaign.  
Then, we would expect 30 English-speaker conversions and 5 Spanish-speaker conversions.\footnote{In this calculation, we assume outcomes are attained through linear interpolation, which we believe to be reasonable in our setting given that we are not in the domain of very small or very large conversion counts. 
}

\subsection*{Survey Methodology}

To understand preferences for SNAP ad budget allocation between English and Spanish speakers (e.g., among the options displayed in Figure \ref{fig:high_pareto}), 
we ran a Qualtrics survey distributed via Prolific to a gender-balanced sample of 2,000 U.S.-based partisans (Republicans or Democrats). Each respondent was randomly assigned to one of five trade-off treatment arms (``high trade-off'', ``low trade-off'', or the three synthetic trade-offs). Survey respondents were introduced to the demographic spectrum in Figure \ref{fig:fairness_compare} and told that Spanish speakers make up 23\% of individuals under the poverty line in San Diego; respondents were also asked several attention checks and comprehension questions to confirm eligibility, yielding 1,532 eligible respondents. Then, respondents were shown a series of pairwise (head-to-head) comparisons of feasible allocation outcomes corresponding to points along the Pareto frontier for their trade-off arm.
For each pairwise comparison, respondents were asked to choose their preferred allocation (e.g., choosing between either Option A or Option B in Figure \ref{fig:fairness_compare}, then choosing between either Option A or Option C, and so on). 

Because we considered six allocation options in each treatment arm, we surveyed respondents about all 15 pairwise comparisons for ease of calculating within-respondent mode for preferred outcome (in the case of ties, the lower allocation share of Spanish speakers was chosen as the ``preferred'' option, meaning that our estimates regarding diversity preferences are conservative).  We apply pairwise trade-off analysis as is common in the marketing literature~\cite{Johnson1976, Salganik2015, green2001} to determine which of the six allocation options (within each treatment arm) is each respondent's most-preferred option. 
This allows us to meet our primary research goal: measuring  the extent to which individuals are willing to exchange some degree of efficiency in order to reach more Spanish speakers in the SNAP online advertising paradigm.  We define ``efficiency'' here as maximizing the number of conversions per dollar, which corresponds to selecting the lowest-possible allocation of Spanish speakers (8\% conversions in the high trade-off arm, and 10\% conversions in the low trade-off arm), since Spanish speakers are, on average, more expensive to target than English speakers.

In addition to pairwise comparisons of allocations, our survey includes a standard slate of demographic questions, as well as two additional questions that 
allow us to further explore individual preferences.

\begin{enumerate}

\item Ideology question:\\
\noindent\fbox{%
    \parbox{22em}{%
    Both Bob and Steve want to help individuals apply for SNAP.  However, they disagree on their approaches towards the comparison questions that you just answered.
    \begin{itemize}
    \item Bob says: ``I wouldn't want the determination of SNAP applicants to take into account people's preferred language.  It's wrong to give preferential treatment to people based on the language they speak, and it is more `fair' to simply allow for the most people (regardless of language) to apply for SNAP.  I always chose the option that led to the most total SNAP applicants.''

    \item Steve says: ``I also don't want any SNAP applicants to get preferential treatment, which is why I preferred for there to be a `fair' number of Spanish-speaking SNAP applicants.  Otherwise, Spanish speakers are disadvantaged relative to English speakers.  So, I always chose the option that had closest to 23\% Spanish speakers.''
    \end{itemize}
    Whose opinion is closer to your own?
    }
}

\item Trolley question (where ($N_1$, $N_2$) is replaced with (1,3) in the low trade-off survey arm, and (1,6) in the high trade-off survey arm):\\
\noindent\fbox{%
    \parbox{22em}{%
        Suppose you must choose between recruiting either $N_1$ Spanish-speaking SNAP applicant, or $N_2$ English-speaking SNAP applicants.  Who do you choose to help?
    }
}
\end{enumerate}

\section*{Results}

\begin{figure*}[ht]
    \centering
    \includegraphics[width=\linewidth]{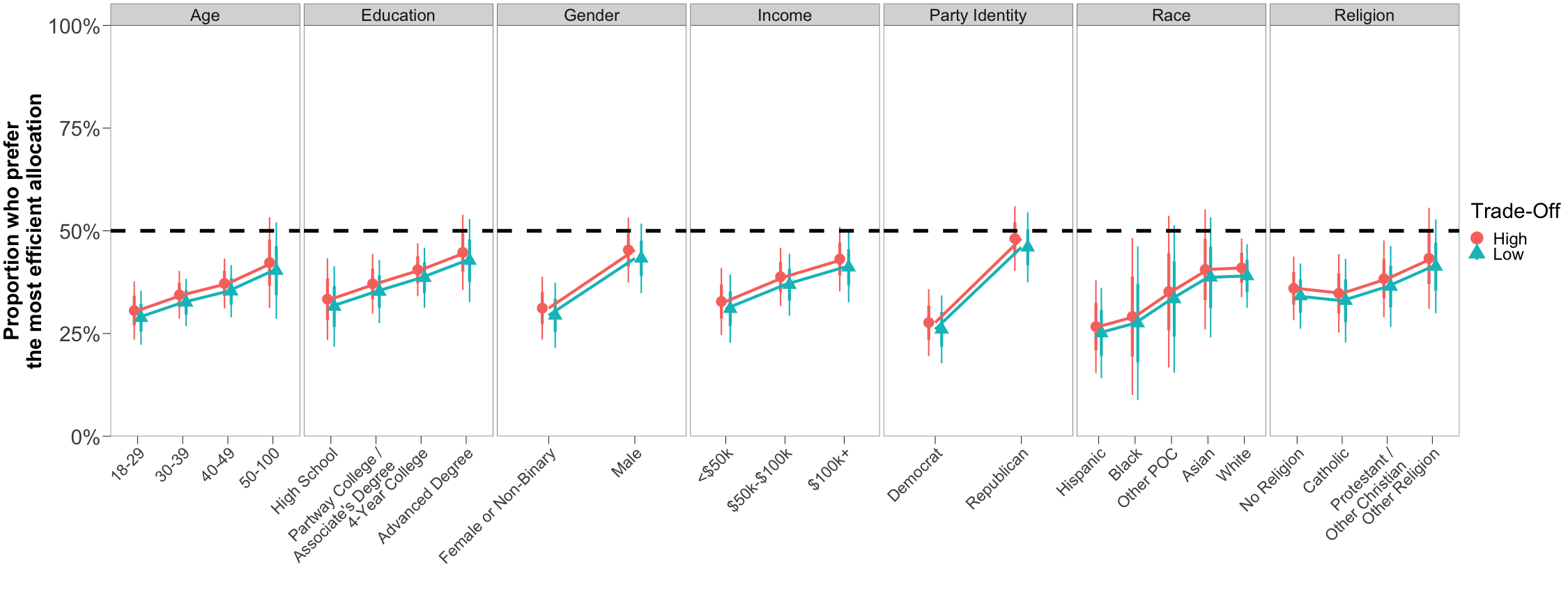}
    \caption{The share of respondents preferring the most efficient allocation (i.e., maximizing conversions by minimizing the number of Spanish speakers to whom ads are presented), as estimated via model-based poststratification. We report 95\% CIs over 1,000 bootstrapped estimates. The black horizontal dashed line represents 50\%; no demographic subgroup has a point estimate surpassing majority preference to maximize conversions.
    The demographic subgroup with the greatest preference for maximizing conversions is self-identified Republicans.}
    \label{fig:ranked_prefs}
\end{figure*}

We find low preference for the most ``efficient'' option across respondent demographic groups, with the largest preference difference being between Republicans and Democrats.  Furthermore, we find stronger support for demographic parity than for maximizing conversions. 

Figure \ref{fig:ranked_prefs} depicts estimates of the share of a demographic group's preference for ``maximizing conversions'' (i.e., the most efficient allocation).
To account for potential differences between our survey sample and the general population, these estimates are generated via model-based poststratification,
with bootstrapped 95\% confidence intervals.
Specifically, using a logistic regression, we first model the preferred options conditional on respondent demographic data on gender, age, party identity, race, religion, education, and income; we then estimate the preferred option on each of 3,840 cells (for each combination of four age groups, four education groups, two gender groups, three income groups, two party identities, five race groups, and four religion groups). 
Finally, we take a weighted average of 
cell-level estimates, with weights equal to the share of the U.S.\ population belonging to each cell based on 2018 AP VoteCast data~\cite{votecast2018,Wang2015}.

\begin{figure}[t]
    \centering
    \includegraphics[width=\linewidth]{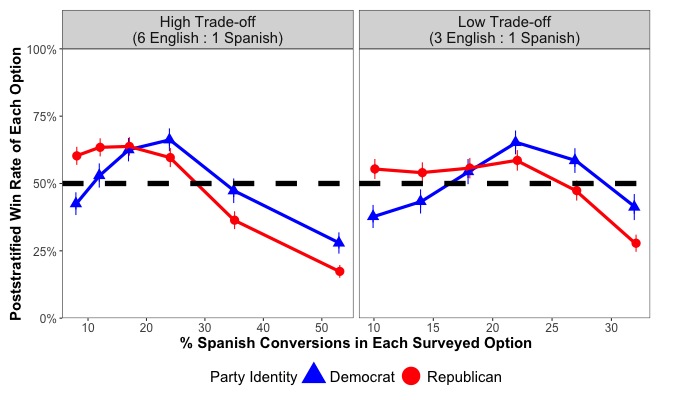}
    \caption{For each of the high and low trade-off survey arms, survey respondent preferences were elicited among six different allocation options (along the x-axis). First, for respondents of a given party identity, we estimate win rates for each of these options, representing the likelihood that an option won a pairwise comparison given that the option was in the running. We then poststratify win rates as done in Figure \ref{fig:ranked_prefs}; in contrast to the preference shares in Figure \ref{fig:ranked_prefs}, here the win rates definitionally yield a mean of 50\% across the six options.  These win rates confirm the Figure \ref{fig:ranked_prefs} finding that the most efficient allocation (i.e., maximizing conversions, the left-most x-axis option) has a higher likelihood of being chosen by Republicans than Democrats. Furthermore, we see a general shift in preferences at the x-axis point nearest the ``demographic parity'' allocation of 23\% Spanish speaker conversions: to the right of this point, less-efficient options are more likely to win among Democrats than Republicans; to the left, the more-efficient options are more likely to win among Republicans than Democrats.}
    \label{fig:poststrat_winrates}
\end{figure}

\begin{figure}[t]
    \centering
    \includegraphics[width=\linewidth]{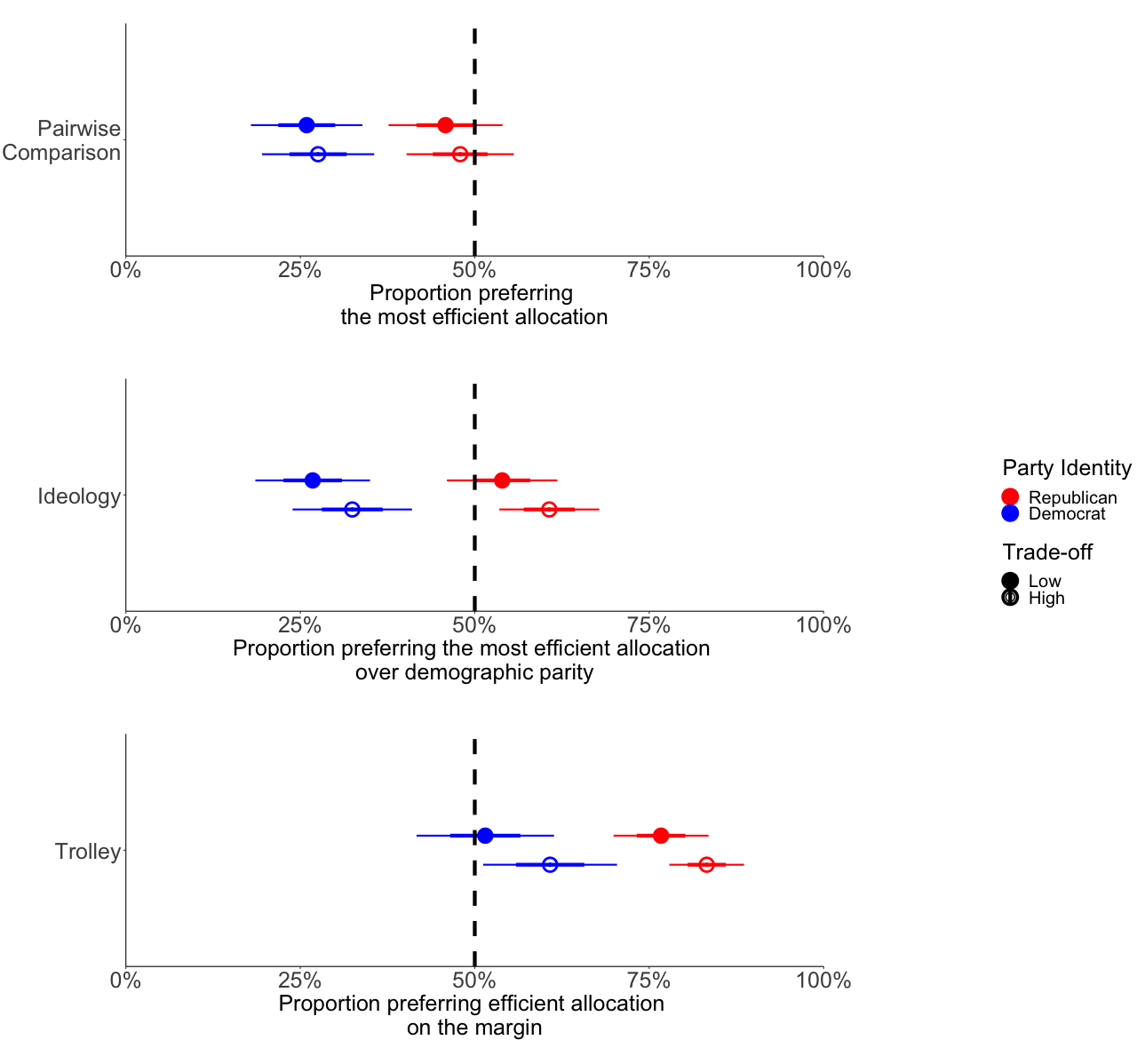}
    \caption{
    Preferences under different elicitation methods; in all panels, red Republican dots are to the right of blue Democrat dots.
    Both the ``ideology'' question (comparing the most efficient allocation to demographic parity) and the ``trolley'' question (eliciting marginal preferences) capture preferences over relatively extreme options. 
    In contrast, pairwise comparison allows respondents to evaluate options that most closely reflect the real trade-offs in the policy problem we consider. 
    }
    \label{fig:ideological}
\end{figure}

Per Figure \ref{fig:ranked_prefs}, for each of the demographic subgroups we consider, 
a minority of individuals in each group (according to our point estimates)
have a preference for maximizing efficiency.
The difference in preference is particularly high between Democrats and Republicans (a 20 percentage-point differential in the high trade-off treatment arm).
Across subgroups, Republicans leaned most toward ``efficiency''---i.e., maximizing conversions---but, even in this case, a majority of Republicans made choices that reflected a desire to balance efficiency with equity. 
In the low trade-off treatment arm, we estimate that 26\% of Democrats and 46\% of Republicans prefer the most-efficient allocation.  In the high trade-off treatment arm, 27\% of Democrats and 48\% of Republicans prefer the most-efficient allocation.

Our results indicate broad support for trading at least some degree of efficiency for reaching more Spanish speakers. One potential confounder is a language-based preference for English speakers, rather than simply preferring efficiency. To test for this, we present Appendix Figure \ref{fig:allslope_multinomial} plotting the same model-based poststratification analysis in Figure \ref{fig:ranked_prefs}, but for synthetic trade-offs where English speakers are instead costlier to convert than Spanish speakers. We find that Republicans remain the demographic subgroup that most prefers to maximize the number of English conversions even when this is no longer the most efficient option---hence, our findings in Figure \ref{fig:ranked_prefs} are conservative and true preference for efficiency (separate to English-language bias) is likely lower.

Next, we seek to understand the extent to which individuals are willing to deviate from the efficient allocation. To understand these more granular preferences, we present poststratified win rates of each allocation option in Figure \ref{fig:poststrat_winrates}. Here, we see that the win rates for the most-efficient allocations (the left-most points) are nearly always far lower than the win rates near ``demographic parity'' of 23\% Spanish speaker conversions. The one exception is Republicans in the high trade-off arm, but for these survey respondents there \emph{still} exist less-efficient allocations with higher win rates than the most-efficient allocation. The ``demographic parity'' allocation has the highest win rate among Democrats in both low and high trade-offs; this peak is less pronounced for Republicans.

In addition to eliciting preferences via pairwise comparison of advertising strategies, we study responses to the two additional survey questions regarding preferences. First, we asked an ``ideological'' question (i.e., the binary choice of ideological alignment with either the most efficient allocation, or the allocation of 23\% Spanish speaker conversions to reflect ``demographic parity'' in the SNAP ads context).  Then, we asked the ``trolley problem'' question (i.e., the binary choice of whether to sacrifice some number of English conversions---dependent on high/low trade-off---to gain one additional Spanish conversion). 

Per Figure \ref{fig:ideological}, we see that the ``ideology'' proportions (measuring preference for the most efficient allocation over the one specific ``demographic parity'' allocation) are 
only slightly higher than the proportion who reveal a preference for efficiency over all other options (as indicated by our analysis of pairwise comparisons).\footnote{Note that the ``ideological'' proportions mirror those from our pairwise comparisons if we subset the win rate analysis from Figure \ref{fig:poststrat_winrates} to only pairwise comparisons between the two options presented in the ideological framing. This yields efficiency-over-demographic-parity preference shares of 31.4--37.3\% for Democrats and 48.6--58\% for Republicans in low-high trade-offs, respectively.}
This result suggests that many individuals not only prefer deviating from the efficient allocation to reach more Spanish speakers, but in fact view demographic parity as more closely capturing their preferences in this context.
That finding is further supported by the pairwise win-rates in Figure \ref{fig:poststrat_winrates}
discussed above.

Next, the trolley problem posed to respondents is a common way to elicit \emph{marginal} preferences (i.e., the extent to which an individual is willing to trade enrolling a single Spanish speaker for some number of English speakers).
In this case, we find a majority of respondents prefer enrolling more English speakers over the lone Spanish speaker---in both the high- and low-trade-off arms, and among both the subset of Democrats and the subset of Republicans.
Critically, though, considering marginal trade-offs only imperfectly 
captures the policy problem at the heart of our application: the key choice is not whether to enroll a single Spanish speaker or multiple English speakers, but rather how to design a recruitment strategy that impacts large numbers of people in both language groups. When presented with options that more closely reflect the trade-offs inherent to our setting, respondents express a stronger preference for incorporating equity considerations into decisions, often selecting an intermediate option---which is not captured by the marginal trade-off question.\footnote{
It is common to assume individuals have linear utilities, which arise naturally if we assume individuals have a fixed value $v_1$ for each conversion of an English speaker and value $v_2$ for each conversion of a Spanish speaker. In this case, marginal utility fully describes one's preferences.
However, if respondents indeed had linear utility functions, then we would expect them to always select an extreme option among the pair-wise choices (i.e., either the most total conversions or most conversions of Spanish speakers). 
We do not see that pattern---many select an intermediate option---which suggests trolley-problem type formulations are ill-suited to eliciting policy preferences in our setting.
}

Finally, we compare our findings to preferences for affirmative action in the workplace and college admissions.  Within our survey, the share of respondents preferring to prioritize Spanish speakers over a purely ``efficient'' advertisting strategy is far higher than those supportive of affirmative action in the workplace (12\% for Republicans and 56\% for Democrats), and who believe race should be a major factor in college admissions (2\% for Republicans and 16\% for Democrats).  
There are many reasons that may explain the preference gap between affirmative action policies and equity in online advertising (e.g., survey question phrasing, the fact that survey respondents may be more directly affected by affirmative action than welfare spending, or that they may be less sensitive to questions about the allocation of a non-profit's funds).
Nonetheless, our findings point towards broader support for equity in machine learning-based algorithmic applications, which has significant implications for future technology policy.

\section*{Discussion \& Limitations}

Our online ad experiments reveal stark disparities in the cost of recruiting English and Spanish speakers into SNAP. Those cost disparities in turn create an inherent tension between efficiency and equity: reaching more Spanish speakers means lowering the overall reach of the advertising campaign, as Spanish speakers are, on average, more expensive to recruit.
To understand popular preferences trading off inclusion, we surveyed a diverse sample of Americans, asking them to choose between different ways to spend the fixed advertising budget. Across groups defined by age, race, gender, education, income, religion, and party identity, we found a majority of respondents in each subgroup preferred sacrificing at least some efficiency to reach more Spanish speakers. 

Our results have immediate implications for the equitable design of algorithms. Deployed optimization algorithms---like those used in online advertising---largely focus on efficiency, but our results reveal that when presented with detailed equity-efficiency trade-offs, the general public would incorporate equity considerations into allocation decisions. 
Further, whereas much of the algorithmic fairness literature has focused on context-agnostic, axiomatic approaches, our results illustrate the value of framing questions of equity in terms of concrete trade-offs that decision-makers, affected communities, and the general public can consider when making difficult choices.

As a result of this research, GetCalFresh has adjusted their Google Ads budget allocations to recruit more Spanish speakers, roughly in line with the demographic parity benchmark (e.g., 23\% Spanish speakers in San Diego County). The task of recruiting SNAP applicants is, of course, not limited to Google Ads. We focus on online recruitment to SNAP via Google Ads due to cost efficiency, but GetCalFresh may additionally expand the budget allocation process to include other platforms (such as Facebook or Bing Ads), and also offline recruitment methods via Community Based Organizations, radio, or text messaging systems. Broadening the scope of our analysis beyond the U.S. welfare system, the efficiency-equity trade-off is also relevant in online services such as online job postings, credit ads, and house listings~\cite{facebook2019}, highlighting the relevance of our work to the design of equitable advertising strategies in these critical applications.

There are several limitations and open questions presented by our study. For example, whose preferences should be elicited to guide policy decisions? To approach this, key questions include: whether decision-makers have perverse incentives, and how much harm could be done to those affected by the decisions~\cite{whittaker2020}.  
It is also particularly important to consider the preferences of the most-affected communities to address historic inequities. We did not have the resources to conduct the survey translated to Spanish and targeted towards SNAP recipients specifically, but doing so would yield more insight into an essential set of perspectives additional to those of our Spanish-speaking and welfare-recipient survey respondents per Appendix Figure \ref{fig:raw_winrates}.
Furthermore, our survey results could be biased due to respondents being motivated to select options that they believed would appeal to the researchers examining their choices, a general lack of incentives to think deeply about the choices at hand, or due to the ``underdog effect'' wherein greater support is given to a smaller-sized group.

Our study shows---at least in the domain we consider---that there is substantial popular support for including equity in algorithm design, even if that means sacrificing some efficiency. This support, perhaps surprisingly, stretches across diverse social and political groups. 
But there is not consensus, and policymakers must still navigate the large gaps between political parties in the extent to which they are willing to trade equity for efficiency.
We hope our approach and results provide a viable path for designing more equitable algorithms.

\appendix
\section{Appendix}

Because certain search keywords are identical in both languages (e.g., ``SNAP''), we must ensure that no cannibalization occurs between the English-targeting campaign and Spanish-targeting campaign, since this would artificially drive up the ad costs (i.e., as the advertiser, we would bid against ourselves).\footnote{Note also that for a variety of reasons (e.g., because Google users may be bilingual, or using a shared household computer with a different language setting than their personal preference, or assisting family members with the SNAP application form), a non-zero number of conversions via the English-targeting Google Ad campaign may fill out the SNAP application in Spanish, and a non-zero number of conversions via the Spanish-targeting Google Ad campaign may fill out the SNAP application in English.}  To avoid this cannibalization between the Spanish-targeting and English targeting campaigns, we run the two campaigns at different times of day in a round robin style. Specifically, we restrict one campaign to run for three 4-hour blocks per day (midnight-4am, 8am-noon, and 4pm-8pm Pacific Time), and we restrict the other campaign to run for the remaining three 4-hour blocks per day (4am-8am, noon-4pm, 8pm-midnight). We randomly pick which campaign begins each week with the first set of 4-hour blocks, and alternate the time block assignment daily (so, the campaign with the midnight-4am slot on Monday would have the 4am-8am slot instead on Tuesday, and so on).  Our experiments are stratified such that weekday and weekend campaign times (the latter having significantly less Google traffic) are roughly even between the English and Spanish-targeting campaigns. For the high trade-off experiment, parameters were: ``Maximize Conversions'' for both language campaigns; English-targeting campaign with a \$385 daily budget, Spanish-targeting campaign with a \$115 daily budget; run from September 28, 2020 to October 12, 2020; conversion counts normalized by total ad budget. For the low trade-off experiment: ``Target CPA'' bidding method set to \$2.97 daily target for both language campaigns; English-targeting campaign with a \$385 daily budget, Spanish-targeting campaign with a \$115 daily budget; run from October 13, 2020 to October 25, 2020; conversion counts normalized by total ad cost.

\begin{table}[t]
\centering
\small\addtolength{\tabcolsep}{-3pt}
\begin{tabular}{lll}
\textbf{Survey treatment arm} & \textbf{\# Democrats} & \textbf{\# Republicans} \\ 
\hline
High trade-off & 161 & 157 \\
Low trade-off & 159 & 149 \\
Equal trade-off & 148 & 153 \\
Flipped low trade-off & 158 & 145 \\
Flipped high trade-off & 140 & 162 \\
\end{tabular}
\caption{Survey respondent counts.}
\end{table}

\begin{table}[!htbp] \centering 
\begin{tabular}{@{\extracolsep{1pt}}lc} 
\\[-1.8ex]\hline 
\hline \\[-1.8ex] 
 & \multicolumn{1}{c}{\textit{Coefficient}} \\ 
\hline \\[-1.8ex] 
 Slope\_High & 0.091 (0.174) \\ 
  Political\_Republican & 0.885$^{***}$ (0.206) \\ 
  Gender\_NotMale & $-$0.497$^{***}$ (0.175) \\ 
  Race\_Hispanic & $-$0.350 (0.305) \\ 
  Race\_Black & $-$0.085 (0.518) \\ 
  Race\_Asian & 0.069 (0.339) \\ 
  Race\_Other\_POC & $-$0.126 (0.459) \\ 
  Religion\_Catholic & $-$0.279 (0.257) \\ 
  Religion\_Other\_Christian & $-$0.183 (0.239) \\ 
  Religion\_Other\_Religion & 0.252 (0.282) \\ 
  Age\_Value & 0.006 (0.006) \\ 
  Education\_Value & 0.060 (0.064) \\ 
  log(Income\_Value) & 0.082 (0.115) \\ 
  Constant & $-$2.118$^{*}$ (1.244) \\ 
 \hline \\[-1.8ex] 
Observations & 625 \\ 
Log Likelihood & $-$390.700 \\ 
Akaike Inf. Crit. & 809.400 \\ 
\hline 
\hline \\[-1.8ex] 
\textit{Note:}  & \multicolumn{1}{r}{$^{*}$p$<$0.1; $^{**}$p$<$0.05; $^{***}$p$<$0.01} \\ 
\end{tabular}
\caption{Low and high trade-off model on binary Pareto-elicited preference for the most efficient allocation.} 
\end{table}

\begin{table}[!htbp] \centering 
\begin{tabular}{@{\extracolsep{1pt}}lc} 
\\[-1.8ex]\hline 
\hline \\[-1.8ex] 
 & \multicolumn{1}{c}{\textit{Coefficient}} \\ 
\hline \\[-1.8ex] 
 Slope\_FlipLow & $-$0.402 (0.247) \\ 
  Slope\_FlipHigh & $-$0.649$^{**}$ (0.253) \\ 
  Political\_Republican & 1.937$^{***}$ (0.292) \\ 
  Gender\_NotMale & $-$0.197 (0.211) \\ 
  Race\_Hispanic & $-$0.618 (0.430) \\ 
  Race\_Black & $-$0.945 (0.794) \\ 
  Race\_Asian & $-$1.970$^{*}$ (1.036) \\ 
  Race\_Other\_POC & $-$0.366 (0.517) \\ 
  Religion\_Catholic & $-$0.125 (0.310) \\ 
  Religion\_Other\_Christian & 0.014 (0.275) \\ 
  Religion\_Other\_Religion & $-$0.278 (0.362) \\ 
  Age\_Value & 0.018$^{**}$ (0.007) \\ 
  Education\_Value & $-$0.206$^{***}$ (0.080) \\ 
  log(Income\_Value) & $-$0.233$^{*}$ (0.135) \\ 
  Constant & 0.196 (1.437) \\ 
 \hline \\[-1.8ex] 
Observations & 904 \\ 
Log Likelihood & $-$307.469 \\ 
Akaike Inf. Crit. & 644.938 \\ 
\hline 
\hline \\[-1.8ex] 
\textit{Note:}  & \multicolumn{1}{r}{$^{*}$p$<$0.1; $^{**}$p$<$0.05; $^{***}$p$<$0.01} \\ 
\end{tabular}
\caption{Synthetic trade-off model on binary Pareto-elicited preference for maximizing English conversions.} 
\end{table} 

\begin{figure}
    \centering
    \includegraphics[width=\linewidth]{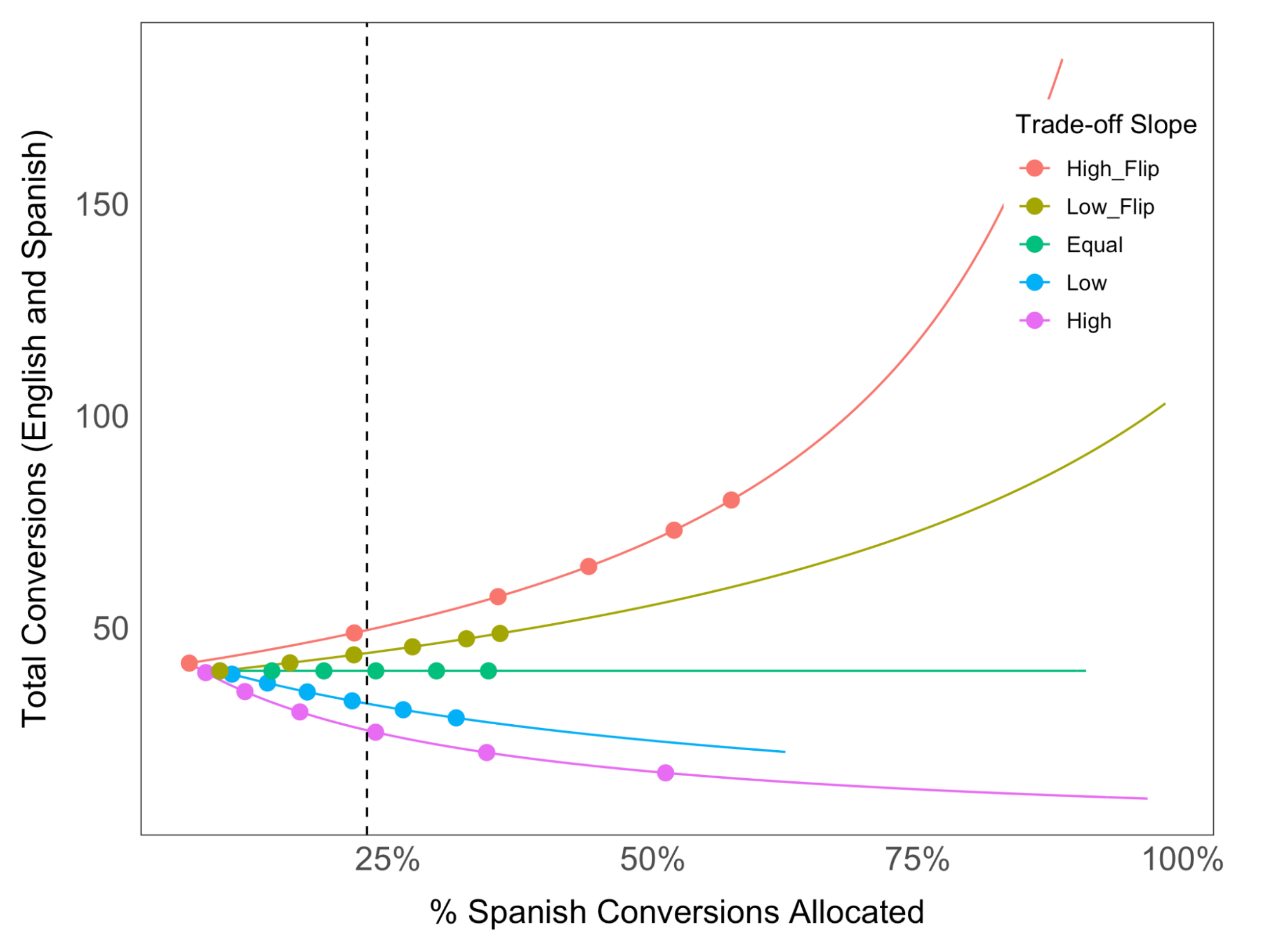}
    \caption{Pareto frontiers generated across both observed and synthetic trade-offs.}
    \label{fig:allslope_pareto}
\end{figure}

\begin{figure*}
    \centering
    \includegraphics[width=\linewidth]{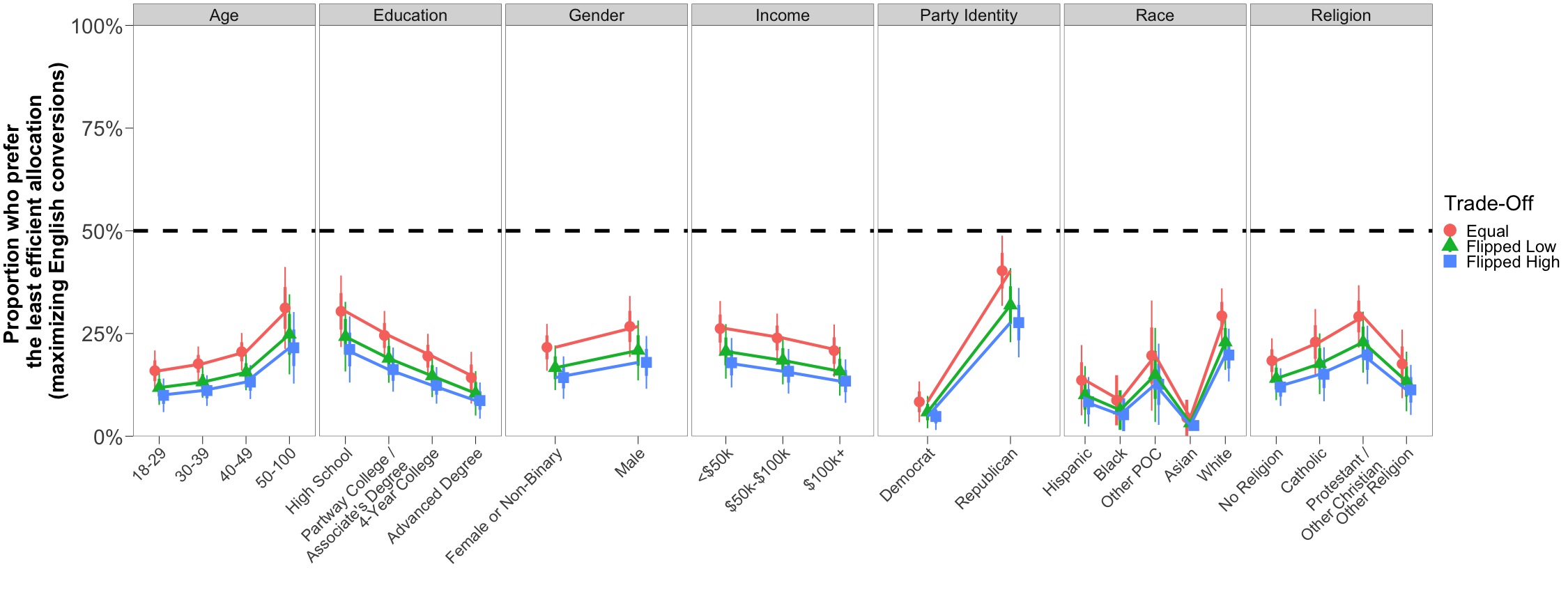}
    \caption{Poststratified estimates of demographic-based preferences in synthetic trade-off slopes wherein the most-efficient option can maximize Spanish conversions (1 English conversion costs 1, 3, or 6 Spanish conversions). The y-axis plots preference for least efficient conversions (i.e., maximizing English conversions and minimizing Spanish conversions---same as the original high and low trade-offs, for which this would be the most efficient conversion). We see higher preference among Republicans (relative to other subgroups) to maximize English conversions in spite of inefficiency, but not majority preference to do so.}
    \label{fig:allslope_multinomial}
\end{figure*}

\begin{figure}[t]
    \centering
    \includegraphics[width=\linewidth]{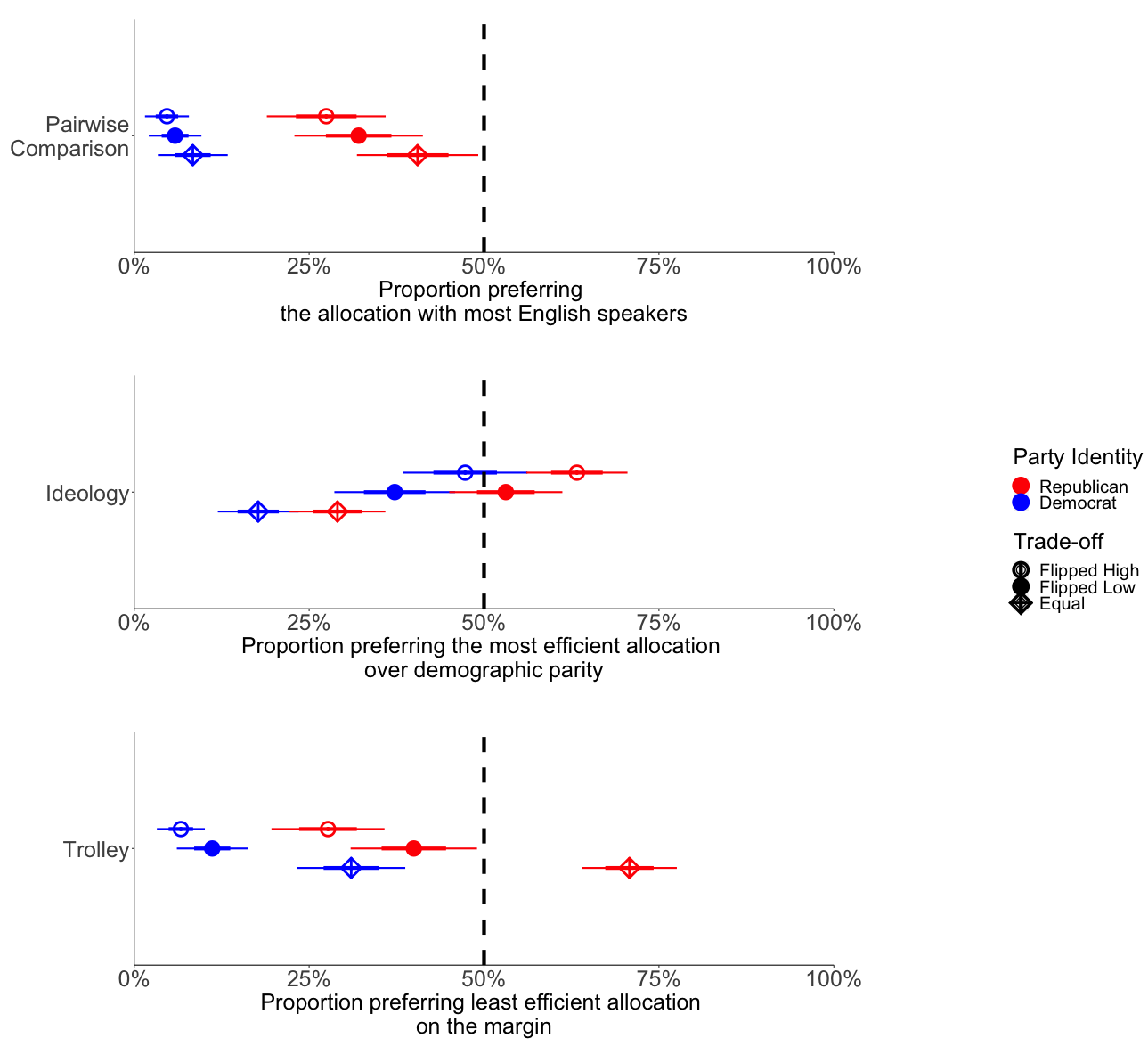}
    \caption{Synthetic trade-off pairwise comparisons and the trolley question display proportions preferring the least-efficient allocation (maximizing English speakers) per Figure \ref{fig:allslope_multinomial}---again, Republicans prefer maximizing English speakers moreso than Democrats, even when this is no longer efficient (in all panels, red Republican dots are to the right of blue Democrat dots). The ideology question now compares the most efficient allocation (maximizing Spanish speaker conversions) to demographic parity (23\% Spanish speakers).}
    \label{fig:ideology_other}
\end{figure}

\begin{figure}[t]
    \centering
    \includegraphics[width=\linewidth]{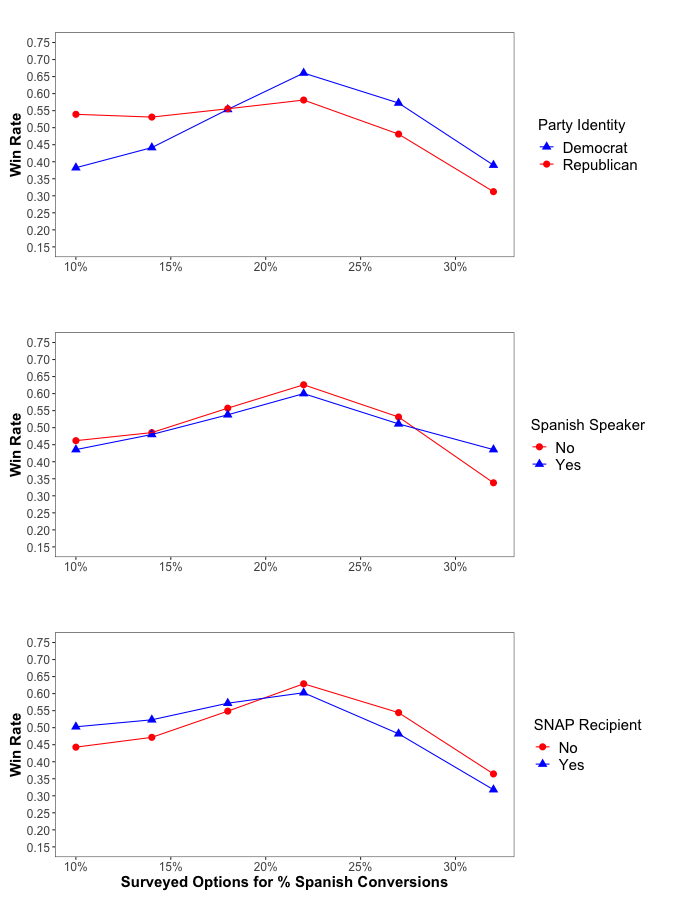}
    \caption{Raw survey win rates (calculated per Figure \ref{fig:poststrat_winrates} but not poststratified) indicate that the differences observed between Republicans and Democrats (top panel) are far larger than the nearly-overlapping lines for survey respondents grouped by: having ever been SNAP recipients (middle panel), or being a Hispanic Spanish speaker (bottom panel). While this does not offer conclusive evidence that preferences do not change by whether one identifies with the relevant community being researched, it does imply that these preference differences are smaller in magnitude than those explained by U.S. political party identity.}
    \label{fig:raw_winrates}
\end{figure}

\section*{Ethical Statement}
While our research aims to generate positive societal impact via increasing equity in SNAP enrollment, there remains a primary ethical consideration: in optimizing for SNAP enrollment among minority demographics, we inherently reduce the SNAP enrollment among majority demographics. Below, we discuss this trade-off and its potential societal impact, as well as concerns arising from data collection. For each source of potential negative societal impact, we describe the principles used to mitigate our concerns.	

The crux of our work involves setting a fixed budget for an advertising bidding algorithm, and optimizing for potential SNAP enrollees who are Spanish speakers. 
For each additional Spanish speaking individual presented with a GetCalFresh ad, we will necessarily decrease the number of English speaking individuals presented with the same ad---potentially by more than one. The long term consequences of our experiment include that certain individuals belonging to majority groups will not be shown the GetCalFresh ad, and will thus have a lower likelihood of filling out GetCalFresh’s SNAP application when searching for the same Google keywords that would otherwise trigger the GetCalFresh ad to be shown. Across California, we hope to see an increase in SNAP applications from individuals who would not otherwise have easily found the online resources to complete the forms, in keeping with GetCalFresh's goal of assisting the neediest individuals.

If applied broadly, our framework can be used to substantiate decision-makers' choices across a range of algorithm-based applications. Depending on the individuals and groups whose preferences are surveyed, this could either yield policy suggestions that propose more equity-based allocations, or ones that propose more efficiency-based allocations as is the norm. The key distinction will stem from \textit{whose} preferences are elicited, and whether their fairness preferences are biased outside the scope of the efficiency-equity trade-off. One way to ameliorate this concern is to ensure representation of underrepresented groups among individuals whose preferences are being elicited~\cite{kasy_2021,whittaker2020}.

As to the ethical challenges of data collection, this research uses data on three fronts: first, from Google ad targeting towards a large swath of Google-users in California; second, from Code for America’s compilation of GetCalFresh applications; and third, from our Prolific survey. In Google ad targeting, we do not have access to any individual-level data; rather, we can only see audience-level statistics (e.g., how many total impressions or clicks were received on an ad).
In the GetCalFresh applications, Code for America continuously tracks all applications that come through its system, but takes particular care to ensure data privacy. For example, even though one could argue the benefits of collecting race-based data from users to optimize for racial equity, the GetCalFresh application does not collect race data at all because the application does not require race information. Further, our team only obtained access to anonymized household-level data pertaining to the research at hand. 
In the Prolific survey, we pre-registered our experiment via \emph{As Predicted} (\#84866), indicating what individual-level data we aimed to collect; we made the decision to not publicly release said data for Prolific user privacy reasons---our dataset includes sensitive information such as political affiliation and income levels, which were imperative to collect to understand the socioeconomic drivers of fairness preferences. Across these three data sources, we have minimized the potential data privacy harm to the extent possible while still allowing for this research to be conducted. The survey question text, along with code reproducing data analysis, are posted on GitHub for reproducibility.

Across both the inherent demographic trade-off and data privacy considerations, we have made the choices we feel best lead to equitable outcomes for the neediest potential GetCalFresh users, with minimal harm to the broader set of potential users. While our research focuses on SNAP within California, there is room for future work both across America, and via analogous food stamp programs globally. However, our design choices would need to be revisited since the concept of neediest recipients may vary considerably in different contexts.

\section{Acknowledgments}
We thank Code for America for their partnership on this project, and Susan Athey for helpful discussions.
\newpage
\bibliography{aaai23}

\end{document}